\documentstyle[preprint,aps,epsf]{revtex}
\tightenlines
\begin{document}

\title{Spin-current modulation and square-wave transmission \\ 
through periodically stubbed electron waveguides}
\author{X. F. Wang$^{a}$, P. Vasilopoulos$^{a}$, and F. M. 
Peeters$^{b}$ \\
\ \\}

\address{ $^{a}$ Concordia University, Department of Physics,\\
 1455 de Maisonneuve Ouest, Montr\'{e}al, Qu\'{e}bec,   H3G  1M8, 
 Canada\\
 \ \\  $^{b}$  Departement Natuurkunde, Universiteit Antwerpen (UIA), \\
Universiteitsplein 1, B-2610, Belgium }

\date{\today} \address{} \address{\mbox{}} \address{\parbox{14cm}{\rm 
\mbox{}\mbox{}\mbox{}
 Ballistic spin transport through waveguides, 
with symmetric or asymmetric double stubs attached to them 
periodically, is studied systematically in the presence of a weak 
spin-orbit coupling that makes the electrons precess.  By an appropriate
choice of the waveguide length and of the stub parameters
injected spin-polarized electrons can be blocked completely and the 
transmission shows a periodic and nearly square-type behavior, with 
values 1 and 0, with wide gaps when only one mode is allowed to 
propagate in the waveguide.  A similar behavior is possible for a 
certain range of the stub parameters even when two-modes can propagate 
in the waveguide and the conductance is doubled.  Such a structure is 
a good candidate for establishing a realistic spin transistor.  A 
further modulation of the spin current can be achieved by inserting
defects in a finite-number stub superlattice. Finite-temperature effects on the
spin conductance are also considered. \\}} \address{\mbox{}} 
\address{\parbox{14cm}{ \rm PACS 72.20.-i, 72.30.+q,73.20.Mf}} 
\maketitle \date{\today} \clearpage

\section{INTRODUCTION}

Spintronics-based quantum computation systems are expected to be one 
of the important successors of the microelectronic-based conventional 
computation systems in the future.  The essential processes of 
realizing spin computation, spin injection into devices, and 
spin-related transport in semiconductors have attracted enthusiastic 
attention in the past few years.  To avoid the practical difficulty of 
integrating devices, some electrical methods, instead of conventional 
ones applying an external magnetic field or employing circularly 
polarized light, are required to induce spin-polarized carriers in 
semiconductor devices.  One simple idea is to use the 
ferromagnet-semiconductor interface to produce spin-polarized 
electrons, but this method must face the mismatch of physical 
parameters between these two quite different materials 
\cite{mole,zhu}.  The employment of diluted magnetic semiconductors 
(DMS), which can match well with other extensively used semiconductors 
like AlGaAs, have provoked a lot of interest in DMS \cite 
{ohno,luci,chan,gurz}.  Recently, based on the Rashba spin-orbit 
interaction, an intrinsic effect in inversely asymmetric or 
asymmetrically confined nanostructures of non-magnetic semiconductors 
(NMS), several designs have been proposed to spin-polarize electronic 
currents in nanostructures \cite {kise,gove}.  This progress in 
spintronics offers the possibility of doing the spin injection in 
conventional materials and is bringing more and more focus on how to 
control and utilize the Rashba effect in these well-known materials 
and well-controlled structures.  The possibility of establishing a 
spin transistor, based on the Rashba interaction, has also been 
considered, but further investigation is required to obtain devices of 
good behavior.

Spin degeneracy of carriers in semiconductors is a result of inversion 
symmetry, in space and time, of the considered system.  By introducing 
a spatial inverse asymmetry, one can realize spin splitting for 
carriers of finite momentum, without applying any external magnetic 
field.  This so-called Rashba spin-orbit interaction \cite{rash,wink} 
has been confirmed experimentally in different semiconductor 
structures \cite {stor,stei,das2,luo}.  In semiconductor 
heterostructures, this spatial inverse asymmetry can be easily 
obtained by either built-in and external electric fields or by the 
position-dependent band edges.  It is found that in many cases, 
especially in narrow gap semiconductor structures, the corresponding 
spin-orbit interaction is a linear function of the electronic momentum 
$\mathbf{k}$\ expressed as the Rashba term $\overrightarrow{\sigma }%
\cdot (\mathbf{k\times E)}$ in the electron Hamiltonian, where $%
\overrightarrow{\sigma }$ is the Pauli spin matrix and $\mathbf{E}$ 
the local electric field.  Thus, a local electric field works on the 
electronic spin like a local magnetic field perpendicular to the 
directions of the electric field and of the electron momentum.  The 
averaged Rashba parameter is proportional to the average electric 
field weighted by the electron probability and can be well controlled 
by a top (back) gate over (below) the device.  Recently, Nitta 
\textit{et al.  }\cite{nitt,enge} studied the dependence of the 
spin-orbit interaction on the surface electric field in an inverted 
InGaAs/InAlAs heterostructure and Grundler \cite{grun} showed that the 
penetration of the electron wave function into the barrier layer can 
greatly enhance the spin-orbit interaction in InAs quantum wells 
because in this case the electrons suffer a stronger effective 
electric field due to the band-edge difference.  The Rashba effect 
began to be considered as one of the powerful tools in making 
spintronic devices after the pioneer proposal of the spin-polarized 
field effect transistor by Datta and Das \cite{datt}.  Mireles and 
Kirczenow \cite{mire} studied in detail the ballistic spin-polarized 
transport and the Rashba spin precession in a semiconductor waveguide 
using a tight-binding model but used somewhat large values for the 
strength of the Rashba parameter.  Based on the fact that the spin 
orientation of electrons in nanostructures depends on the direction of 
their momentum, Kiselev and Kim\cite{kise} proposed a T-shaped spin
filter while Governale et al.\cite{gove} announced recently the 
possibility of making a more effective spin filter with tunnel-coupled 
electron waveguides.  Spin injection into non-magnetic semiconductors 
with the help of magnetic metals \cite{zhu,kire} or diluted magnetic 
semiconductors \cite {gurz} has also been intensively studied.  
However, it is not yet known how to effectively control the 
spin-polarized flux in those ballistic transport devices and 
waveguides, though ballistic electronic transport, disregarding spin 
polarization, has been studied in detail in the past several years 
\cite {debr,tak1,peet,sols,tak2}.  In a previous paper \cite{wang}, we 
showed briefly how spin-polarized transport can be produced and 
controlled in stubbed waveguides when only spin-up or spin-down 
electrons are injected and only one mode is propagating in the 
waveguide.

In the present paper we study in detail the possibility of a spin 
transistor which can control the flux strength (transmission rate) and 
spin orientation using periodically stubbed semiconductor waveguides 
in which the Rashba effect is present.  In addition, we consider the 
case when two modes are allowed to propagate in the waveguide as well 
as that of ''defects'' in a periodic array of stubbed waveguides.  We 
further consider  the case of injected electrons polarized partially up 
and partially down and new stub shapes
as well as the influence of finite temperatures on the conductance.  
The results obtained are 
mentioned in the abstract and are detailed as follows.  In Sec.  II we 
present the formalism and in Sec.  III the numerical results for one 
(III.A) and two propagating modes (III.B) with spin-up injection.  
Results for finite superlattices with defects are considered in Sec.  
III.C and the injection of electrons with their spins polarized in an 
arbitrary direction in Sec.  III.D. Conclusions follow in Sec.  IV.

\section{FORMALISM}

For a typical two-dimensional (2D) electronic system in the $x-y$ 
plane in narrow gap semiconductor nanostructures such as InGaAs/InAlAs 
quantum wells, the one-electron Hamiltonian including the lowest order 
of spin-orbit interaction can be expressed as

\begin{eqnarray}
H^{2D} &=&\frac{\vec{p}^{2}}{2m^{\ast }}+\frac{\alpha }{\hbar 
}(\vec{\sigma}%
\times \vec{p})_{z}=-\frac{\hbar ^{2}}{2m^{\ast 
}}\vec{\nabla}^{2}+i\alpha
(\sigma _{y}
\frac{\partial}{\partial x}
-\sigma _{x}
\frac{\partial}{\partial y}
)  \nonumber \\
&=&\left[ 
\begin{array}{cc}
-\frac{\hbar ^{2}}{2m^{\ast }}\vec{\nabla}^{2} &
\alpha\nabla^-\\
-\alpha\nabla^+ &
-\frac{\hbar ^{2}}{2m^{\ast }}\vec{\nabla}^{2}
\end{array}
\right],
\end{eqnarray}
where $\vec{\nabla}^{2}=\partial ^{2}/\partial 
x^{2}+\partial ^{2}/\partial y^{2}$, and
$\nabla^{\pm}= \partial/\partial x\pm i\partial/\partial y$.
The parameter $\alpha $
measures the \noindent strength of the spin-orbit coupling and is 
proportional to the interface electric field; $\vec{\sigma}=(\sigma 
_{x},\sigma _{y},\sigma _{z})$ denotes the spin Pauli matrices, and 
${\vec{p}}$ is the momentum operator.  In the presence of $\alpha $ we 
assume that the new wave function has the form

\begin{equation}
\Psi (k_{x},k_{y})=e^{ik_{x}x+ik_{y}y}\sum_{\sigma }C^{\sigma }|\sigma
\rangle =e^{ik_{x}x+ik_{y}y}\left( 
\begin{array}{c}
C^{+} \\ 
C^{-}
\end{array}
\right),
\end{equation}
with $|\sigma \rangle ={\tiny\left(
\array{c}
1 \\ 
0
\endarray
\right)}$ (spin up) or ${\tiny \left(
\array{c}
0 \\ 
1
\endarray
\right)}$ (spin down). The solutions of the equation $H^{2D}\Psi
(k_{x},k_{y})=E\Psi (k_{x},k_{y})$ are readily obtained as 
\[
\Psi ^{\pm }(k_{x},k_{y})=\frac{1}{\sqrt{2}}\left( 
\begin{array}{c}
1 \\ 
\frac{\pm k_{y}\mp ik_{x}}{k}
\end{array}
\right) ; 
\]
the corresponding eigenvalues are
\begin{equation}
E^{\pm }(k_{x},k_{y})=\frac{\hbar ^{2}}{2m^{\ast }}k^{2}\pm \alpha k,
\end{equation}
where $k=\sqrt{k_{y}^{2}+k_{x}^{2}\text{.}}$ The electrons are now spin 
polarized and oriented perpendicular to the electronic momentum in the 
2D plane.

\begin{figure}[tpb]
\begin{center}
\addvspace{0 cm}
\leavevmode
\epsfysize=150pt
\epsfxsize=265pt
\epsffile{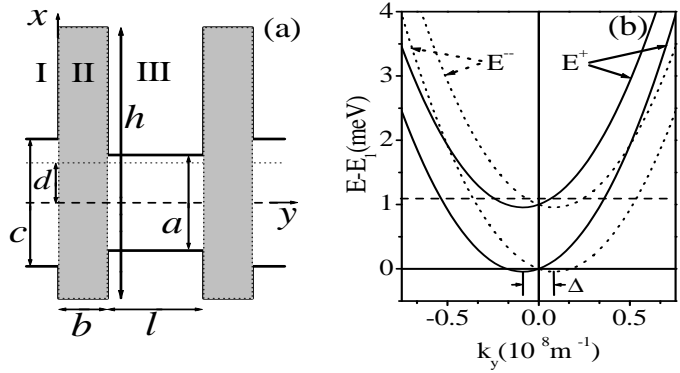}
\end{center}
\caption{
\label{fig1}(a) Schematics of a stub tuner with two units. The waveguide segment
between the two stubs (shaded regions of width $b$) has length $l$ and 
widths $a$ and $c$ as indicated.  The asymmetry parameter $d$ is the distance
between the lines, along the $y$ axis, that bisect the main waveguide 
and the stubs as indicated.
(b) 
Dispersion relation for a waveguide based on Eq. (\ref{disp}). The
horizontal dashed line indicates the four possible values of the wavector
for the same energy $E$.}
\end{figure}

If the electron gas is confined along the $x$ direction by a potential $V(x)$, 
such as the one in the stubbed waveguide shown in Fig. \ref{fig1}(a), we have a
quasi-one-dimensional (Q1D) electronic system. The Hamiltonian 
becomes 

\begin{equation}
H^{\text{Q1D}}=\left[ 
\begin{array}{cc}
-\frac{\hbar ^{2}}{2m^{\ast }}\vec{\nabla}^{2}+V(x) &
\alpha\nabla^-\\
-\alpha\nabla^+ &
-\frac{\hbar ^{2}}{2m^{\ast }}\vec{\nabla}^{2}+V(x)
\end{array}
\right].
\end{equation}
Denoting  by $\phi _{n}(x)$ the solutions of the equation
$\left[ -(\hbar ^{2}/2m^{\ast })\partial ^{2}/\partial
x^{2}+V(x)\right] \phi _{n}(x)=E_{n}\phi _{n}(x)$, we can express the 
Q1D eigenstates in the form 

\begin{equation}
\Psi _{k_{y}}(x,y)=e^{ik_{y}y}\sum_{n,\sigma }\phi _{n}(x)C_{n,\sigma
}|\sigma \rangle =e^{ik_{y}y}\sum_{n}\phi _{n}(x)\left( 
\begin{array}{c}
C_{n}^{+} \\ 
C_{n}^{-}
\end{array}
\right);
\end{equation}
in each of the regions I, II, or III we have $\phi _{n}(x)=\sin (n\pi
 (x+w/2)/w)$, where $w$ is the width of the region along $x$. 
Then the equation $H^{\text{Q1D}}\Psi =E\Psi $ takes the form 
\begin{equation}
\sum_{n}\left( 
\begin{array}{c}
\lbrack E_{n}+\hbar ^{2}k_{y}^{2}/2m^{\ast }-E]\phi 
_{n}C_{n}^{+}+\alpha
k_{x}\phi _{n}C_{n}^{-}+\alpha \phi _{n}^{\prime }C_{n}^{-} \\ 
\alpha k_{y}\phi _{n}C_{n}^{+}-\alpha \phi _{n}^{\prime
}C_{n}^{+}+[E_{n}+\hbar ^{2}k_{y}^{2}/2m^{\ast }-E]\phi _{n}C_{n}^{-}
\end{array}
\right) =0,
\end{equation}
where $\phi_{n}^{\prime}=d\phi/dx$. 

Mutiplying both sides with $\phi _{m}(x)$ and integrating over $x$ 
leads to (%
$\int dx\phi _{m}(x)\phi _{n}(x)=\delta _{mn}$)

\begin{equation}
\left( 
\begin{array}{c}
\lbrack E_{m}+\hbar ^{2}k_{y}^{2}/2m^{\ast }-E]C_{m}^{+}+\alpha
k_{x}C_{m}^{-}+\sum_{n}\int dx\alpha \phi _{m}\phi _{n}^{\prime 
}C_{n}^{-}
\\ 
\alpha k_{x}C_{m}^{+}-\sum_{n}\int dx\alpha \phi _{m}\phi _{n}^{\prime
}C_{n}^{+}+[E_{m}+\hbar ^{2}k_{y}^{2}/2m^{\ast }-E]C_{m}^{-}
\end{array}
\right) =0.  \label{crosse}
\end{equation}
According to degenerate perturbation theory, if the inequality 
\[
\left| \frac{\left( H_{so}\right) _{nm}}
{E_{m}-E_{n}}\right| =\left| \frac{\alpha \int dx\phi _{m}\phi 
_{n}^{\prime }%
}{E_{m}-E_{n}}\right| \ll 1 
\]
holds, we can neglect the subband mixing term $\int dx\phi _{m}\phi
_{n}^{\prime }$. Then Eq. (\ref{crosse}) becomes

\begin{equation}
\left[ 
\begin{array}{cc}
E_{m}+(\hbar ^{2}/2m^{\ast })k_{y}^{2}-E & \alpha k_{y} \\ 
\alpha k_{y} & E_{m}+(\hbar ^{2}/2m^{\ast })k_{y}^{2}-E
\end{array}
\right] \left( 
\begin{array}{c}
C_{m}^{+} \\ 
C_{m}^{-}
\end{array}
\right) =0
\end{equation}
and its eigenvalues are

\begin{equation}
E^{\pm }(k_{y})=E_{m}+(\hbar ^{2}/2m^{\ast })k_{y}^{2}\pm \alpha 
k_{y}.
\label{disp}
\end{equation}
The eigenvectors corresponding to $E^{+},E^{-}$ satisfy $C_{m}^{+}=\pm
C_{m}^{-}$. Accordingly, the spin eigenfunctions are taken as
$|\pm \rangle =
\frac{1}{\sqrt{2}} 
{\tiny\left(
\array{c}
1\\
\pm 1 \endarray \right)}$.  
The dispersion relation $E^{\pm }(k_{y})$ 
versus $k_{y}$ resulting from Eq.  (\ref{disp}) is shown in Fig.  
\ref{fig1}(b).  Note that the minimum of subbands are shifted from
$k_{y}=0$ by the value $\Delta /2=m^{\ast }\alpha /\hbar ^{2}$.  In 
each subband, electrons of the same energy $E$ have four different
 momentum values $k_{y}$, i.e., the positive or negative 
$k_{y}^{+}$ and $k_{y}^{-}$ values pertaining to the branches 
$E^{+}(k_{y})$ and $E^{-}(k_{y})$.  In fact, electrons belonging to 
different branches have opposite spin orientation in the $x$ direction.  
The difference in wave vectors $k_{y}^{+}$ and $k_{y}^{-}$, resulting 
from $ E^{+}=E^{-}=E$, reads

\begin{equation}
k_{y}^{-}-k_{y}^{+}=2m^{\ast }\alpha /\hbar ^{2}=\Delta.
\label{delt}
\end{equation}

Now let us consider the transmission process when an electron of
energy $E$
is incident from the left to a stubbed waveguide as illustrated in 
Fig. \ref
{fig1}(a). The electron wavefunction is decomposed into the plus
${\tiny\left(
\array{c}
1 \\ 
1
\endarray
\right)}$ and minus ${\tiny\left(
\array{c}
1 \\ 
 -1 \endarray \right)}$
  branches and the procedure outlined above 
 applies to each of the three regions labelled I, II, and III in Fig.  
 \ref{fig1}(a).  In each region we have $\phi _{n}(x)=\sin (n\pi
 (x+w/2)/w)$, where $w$ is the width of the region along $x$.  
 Including spin and referring to Fig.  \ref{fig1}(b) we can write the
 eigenfunction of energy $E$ in region I as

\begin{eqnarray}
\phi _{1} &=&\sum_{m}\left\{ a_{1m}^{+}e^{i\beta _{m}y}\left( 
\begin{array}{c}
1 \\ 
1
\end{array}
\right) +a_{1m}^{-}e^{i(\beta _{m}+\Delta )y}\left( 
\begin{array}{c}
1 \\ 
-1
\end{array}
\right) \right.  \nonumber \\
&&+\left. b_{1m}^{+}e^{-i(\beta _{m}+\Delta )y}\left( 
\begin{array}{c}
1 \\ 
1
\end{array}
\right) +b_{1m}^{-}e^{-i\beta _{m}y}\left( 
\begin{array}{c}
1 \\ 
-1
\end{array}
\right) \right\}  \nonumber \\
&\times &\sin (c_{m}(x+c/2))
\end{eqnarray}
and in region III as 
\begin{eqnarray}
\phi _{2} &=&\sum_{n}\left\{ a_{2n}^{+}e^{i\alpha _{n}(y-b)}\left( 
\begin{array}{c}
1 \\ 
1
\end{array}
\right) \right. +a_{2n}^{-}e^{i(\alpha _{n}+\Delta )(y-b)}\left( 
\begin{array}{c}
1 \\ 
-1
\end{array}
\right)  \nonumber \\
&+&b_{2n}^{+}e^{-i(\alpha _{n}+\Delta )(y-b)}\left( 
\begin{array}{c}
1 \\ 
1
\end{array}
\right) +\left. b_{2n}^{-}e^{-i\alpha _{n}(y-b)}\left( 
\begin{array}{c}
1 \\ 
-1
\end{array}
\right) \right\}  \nonumber \\
&\times &\sin (a_{n}(x+a/2));
\end{eqnarray}
here $c_{m}=m\pi /c$, $\beta _{m}=(2m^{\ast }E-c_{m}^{2})^{1/2}$, $%
a_{n}=n\pi /a$, and $\alpha _{n}=(2m^{\ast }E-a_{n}^{2})^{1/2}$ where 
$%
m,n=1,2,3,\cdots $ denote the order of the transverse modes.  The 
symbols
$a_{1m}^{\sigma }$, $b_{1m}^{\sigma }$, $a_{2n}^{\sigma }$,
$b_{2n}^{\sigma }$ ($\sigma =\pm $) represent the coefficients of 
different electronic modes existing in the device.  Similar to the 
procedure of matching electronic wavefunction when spin is disregarded 
\cite{tak1}, in region II we use two auxiliary sets of solutions to 
the wave equation, one of which matches the wire on the left and the 
other on the right, with each vanishing elsewhere on the boundary: 
$\phi _{s}=\sum_{k}(\chi _{k}^{L}+\chi _{k}^{R})$.  The appropriate 
boundary conditions are

\begin{eqnarray}
\chi _{k}^{R}(y &=&0,x)=0;  \nonumber \\
\chi _{k}^{R}(y &=&b,x)=\left\{ 
\begin{array}{cc}
0, & x<-a/2; \\ 
\sin [a_{k}(x+a/2)]\left( 
\begin{array}{c}
a_{2k}^{+}+a_{2k}^{-}+b_{2k}^{+}+b_{2k}^{-} \\ 
a_{2k}^{+}+a_{2k}^{-}-b_{2k}^{+}-b_{2k}^{-}
\end{array}
\right), & -a/2<x<a/2; \\
0, & x>a/2;
\end{array}
\right.  \label{right}
\end{eqnarray}
\begin{equation}
\chi _{k}^{L}(y=0,x)=\left\{ 
\begin{array}{cc}
0, & x<-c/2; \\ 
\sin [c_{k}(x+c/2)]\left( 
\begin{array}{c}
a_{1k}^{+}+a_{1k}^{-}+b_{1k}^{+}+b_{1k}^{-} \\ 
a_{1k}^{+}+a_{1k}^{-}-b_{1k}^{+}-b_{1k}^{-}
\end{array}
\right), & -c/2<x<c/2; \\
0, & x>c/2;
\end{array}
\right.  \nonumber
\end{equation}
and $\chi _{k}^{L}(y=b,x)=0$. $\chi _{k}^{R}$ can be expanded as

\begin{eqnarray}
\chi _{k}^{R} &=&\sum_{n}\left\{ u_{n}^{R}e^{i\gamma _{n}y}\left( 
\begin{array}{c}
1 \\ 
1
\end{array}
\right) \right. +v_{n}^{R}e^{i(\gamma _{n}+\Delta )y}\left( 
\begin{array}{c}
1 \\ 
-1
\end{array}
\right)  \nonumber \\
&-&u_{n}^{R}e^{-i(\gamma _{n}+\Delta )y}\left( 
\begin{array}{c}
1 \\ 
1
\end{array}
\right) -\left. v_{n}^{R}e^{-i\gamma _{n}y}\left( 
\begin{array}{c}
1 \\ 
-1
\end{array}
\right) \right\}  \nonumber \\
&\times &\sin (h_{n}(x+h/2-d)).  \label{rightw}
\end{eqnarray}
Multiplying Eq. (\ref{right}) by $\sin (h_{m}(x+h/2-d))$, integrating
over $x$ ($-a/2\leq x\leq a/2$) and using Eq. (\ref{rightw}), we obtain 

\begin{equation}
  u_{m}^{R}=2(a_{2k}^{+}+b_{2k}^{+})I_{km}^{R}/K_{m}^{+}, \ \ \ 
v_{m}^{R}=-2(a_{2k}^{-}+b_{2k}^{-})I_{km}^{R}/K_{m}^{-} 
\label{uvr}
\end{equation}
where 

\begin{equation}
I_{km}^{R}=\int_{-a/2}^{a/2}dx\sin [a_{k}(x+a/2)]\sin 
(h_{m}(x+h/2-d), \ \  K_{m}^{\pm }=h[e^{\pm i\gamma _{m}b}-e^{\mp 
i(\gamma _{m}+\Delta )b}], 
\label{ikr}
\end{equation}
and $ \gamma _{m}=(2m^{\ast }E-h_{m}^{2})^{1/2}$.
Similarly, $\chi _{k}^{L}$ can be expanded as

\begin{eqnarray}
\chi _{k}^{L} &=&\sum_{n}\left\{ u_{n}^{L}e^{i\gamma _{n}(y-b)}\left( 
\begin{array}{c}
1 \\ 
1
\end{array}
\right) \right. +v_{n}^{L}e^{i(\gamma _{n}+\Delta )(y-b)}\left( 
\begin{array}{c}
1 \\ 
-1
\end{array}
\right)  \nonumber \\
&-&u_{n}^{L}e^{-i(\gamma _{n}+\Delta )(y-b)}\left( 
\begin{array}{c}
1 \\ 
1
\end{array}
\right) -\left. v_{n}^{L}e^{-i\gamma _{n}(y-b)}\left( 
\begin{array}{c}
1 \\ 
-1
\end{array}
\right) \right\}  \nonumber \\
&\times &\sin (h_{n}(x+h/2-d))
\end{eqnarray}
where 

\begin{equation}
u_{m}^{L}=2(a_{1k}^{+}+b_{1k}^{+})I_{km}^{L}/K_{m}^{-}, \ \ \ 
v_{m}^{L}=-2(a_{1k}^{-}+b_{1k}^{-})I_{km}^{L}/K_{m}^{+}\label{uvl}
\end{equation}
 and  

\begin{equation}
I_{km}^{L}=\int_{-c/2}^{c/2}dx\sin [c_{k}(x+c/2)]\sin 
[h_{m}(x+h/2-d)]
\label{ikl}
\end{equation}

Requiring the continuity of the derivative of the wavefunction at the
interfaces, $\phi _{1}$ and $\phi _{2}$ must satisfy the following 
equation: 
\begin{equation}
\left[ 
\begin{array}{cccc}
\widehat{B}+\widehat{\beta } & \widehat{B}-\widehat{\beta }^{\prime } 
& 0 & 0
\\ 
\widehat{F} & \widehat{F} & 0 & 0 \\ 
0 & 0 & \widehat{D}-\widehat{\beta } & \widehat{D}+\widehat{\beta 
}^{\prime }
\\ 
0 & 0 & \widehat{H} & \widehat{H}
\end{array}
\right] \left( 
\begin{array}{c}
\widehat{a}_{1}^{+} \\ 
\widehat{b}_{1}^{+} \\ 
\widehat{a}_{1}^{-} \\ 
\widehat{b}_{1}^{-}
\end{array}
\right) =\left[ 
\begin{array}{cccc}
\widehat{A} & \widehat{A} & 0 & 0 \\ 
\widehat{E}-\widehat{\alpha } & \widehat{E}+\widehat{\alpha }^{\prime 
} & 0
& 0 \\ 
0 & 0 & \widehat{C} & \widehat{C} \\ 
0 & 0 & \widehat{G}+\widehat{\alpha } & \widehat{G}-\widehat{\alpha }%
^{\prime }
\end{array}
\right] \left( 
\begin{array}{c}
\widehat{a}_{2}^{+} \\ 
\widehat{b}_{2}^{+} \\ 
\widehat{a}_{2}^{-} \\ 
\widehat{b}_{2}^{-}
\end{array}
\right).
\end{equation}
The elements of sub-matrices $\widehat{A}$, $\widehat{B}$, 
$\widehat{C}$, $%
\widehat{D}$, $\widehat{E}$, $\widehat{F}$, $\widehat{G}$, 
$\widehat{H}$, $%
\widehat{\alpha }$, $\widehat{\beta }$, $\widehat{\alpha }^{\prime 
}$, $%
\widehat{\beta }^{\prime }$ have values 
\begin{eqnarray}
A_{lk} &=&\sum_{m}4(2\gamma _{m}+\Delta 
)I_{lm}^{L}I_{km}^{R}/(cK_{m}^{+}),
\\
B_{lk} &=&\sum_{m}4[\gamma _{m}e^{-i\gamma _{m}b}+(\gamma _{m}+\Delta
)e^{i(\gamma _{m}+\Delta )b}]I_{lm}^{L}I_{km}^{L}/(cK_{m}^{-}), \\
C_{lk} &=&\sum_{m}-4(2\gamma _{m}+\Delta 
)I_{lm}^{L}I_{km}^{R}/(cK_{m}^{-}),
\\
D_{lk} &=&\sum_{m}-4[\gamma _{m}e^{i\gamma _{m}b}+(\gamma _{m}+\Delta
)e^{-i(\gamma _{m}+\Delta )b}]I_{lm}^{L}I_{km}^{L}/(cK_{m}^{+}), \\
E_{lk} &=&\sum_{m}4[\gamma _{m}e^{i\gamma _{m}b}+(\gamma _{m}+\Delta
)e^{-i(\gamma _{m}+\Delta )b}]I_{lm}^{R}I_{km}^{R}/(aK_{m}^{+}), \\
F_{lk} &=&\sum_{m}4(2\gamma _{m}+\Delta 
)I_{lm}^{R}I_{km}^{L}/(aK_{m}^{-}),
\\
G_{lk} &=&\sum_{m}-4[\gamma _{m}e^{-i\gamma _{m}b}+(\gamma _{m}+\Delta
)e^{i(\gamma _{m}+\Delta )b}]I_{lm}^{R}I_{km}^{R}/(aK_{m}^{-}), \\
H_{lk} &=&\sum_{m}-4(2\gamma _{m}+\Delta 
)I_{lm}^{R}I_{km}^{L}/(aK_{m}^{+}),
\end{eqnarray}
where $\beta _{lk}=\beta _{k}\delta _{lk}$, $\alpha _{lk}=\alpha _{k}\delta 
_{lk}$%
, $\beta _{lk}^{\prime }=(\beta _{k}+\Delta )\delta _{lk}$, and $\alpha
_{lk}^{\prime }=(\alpha _{k}+\Delta )\delta _{lk}$.

Together with the matrix $\widehat{P}$ corresponding to a waveguide 
segment
of length $l$ rather than to a stub, 
\begin{equation}
\left( 
\begin{array}{c}
\widehat{a}_{1}^{+} \\ 
\widehat{b}_{1}^{+} \\ 
\widehat{a}_{1}^{-} \\ 
\widehat{b}_{1}^{-}
\end{array}
\right) =\widehat{P}\left( 
\begin{array}{c}
\widehat{a}_{2}^{+} \\ 
\widehat{b}_{2}^{+} \\ 
\widehat{a}_{2}^{-} \\ 
\widehat{b}_{2}^{-}
\end{array}
\right) =\left[ 
\begin{array}{cccc}
e^{-i\alpha _{m}l} & 0 & 0 & 0 \\ 
0 & e^{i(\alpha _{m}+\Delta )l} & 0 & 0 \\ 
0 & 0 & e^{-i(\alpha _{m}+\Delta )l} & 0 \\ 
0 & 0 & 0 & e^{i\alpha _{m}l}
\end{array}
\right] \left( 
\begin{array}{c}
\widehat{a}_{2}^{+} \\ 
\widehat{b}_{2}^{+} \\ 
\widehat{a}_{2}^{-} \\ 
\widehat{b}_{2}^{-}
\end{array}
\right), \label{mwg}
\end{equation}
we can connect the incident waves (to the left of region I) with the
outgoing ones (to the right of region III) via a transfer matrix 
$\hat{M}$

\begin{equation}
\left( 
\begin{array}{c}
\widehat{a}_{in}^{+} \\ 
\widehat{b}_{in}^{+} \\ 
\widehat{a}_{in}^{-} \\ 
\widehat{b}_{in}^{-}
\end{array}
\right) =\hat{M}\left( 
\begin{array}{c}
\widehat{a}_{out}^{+} \\ 
\widehat{b}_{out}^{+} \\ 
\widehat{a}_{out}^{-} \\ 
\widehat{b}_{out}^{-}
\end{array}
\right).
\end{equation}
Here $\hat{M}$ is a $4\times 4$ matrix with sub-matrices 
$\hat{M}_{mn}$ as its elements.  If we assume that
a spin-coherent electron 
beam of single energy is injected into the device from the left and it 
is detected at its right end by an analyzer, there is no 
backward propagating wave at the output and 
$\widehat{b}_{out}^{+}=\widehat{b}_{out}^{-}=0$.  The coefficients of 
spin-up and spin-down electrons\ are 
$\widehat{a}_{in}^{+}+\widehat{a}%
_{in}^{-}$ and $\widehat{a}_{in}^{+}-\widehat{a}_{in}^{-}$ for the 
input wave and $\widehat{a}_{out}^{+}+\widehat{a}_{out}^{-}$ and 
$\widehat{a}_{out}^{+}-%
\widehat{a}_{out}^{-}$ for the output wave, respectively, if we neglect 
the injection mismatch of electronic momentum between different 
electron branches.  Then the spin-dependent transmission rate can be 
calculated similar to the procedure used when the spins are 
disregarded.

In the design of a spin transistor first  proposed by Datta and Das, we 
can introduce the above stubbed waveguide together with the Rashba 
effect into the device to control and flip the spin current.  To see 
this effect clearly, we connect a spin polarizer (analyzer) to the 
left (right) of the structure and inject spin-up polarized electrons 
into it and detect the polarization of the outgoing electrons.  The 
electron beam is equally decomposed into the plus branch
 ${\tiny\left( 
\array{c} 1 \\
1
\endarray
\right)}$ and the minus branch
${\tiny\left( 
\array{c}
1 \\ 
-1 \endarray \right)}$ 
if the momentum difference between two branches 
can be neglected when considering the wavefunction match of input and 
output.  Since the electrons belong to different branches will get 
different phases when propagating along the device, the output spin 
orientation of the electrons, which is a result of the composition of 
electronic wavefunction of the two branches when leaving from the 
waveguide, can be different from the initial one.  In our case, the 
spin-up transmission rate $T^{+}$ and spin-down transmission rate 
$T^{-}$\ read
\begin{equation}
T^{\pm }=
\sum_{mn}T^{\pm}_{mn}=
\sum_{mn}\frac{\alpha _{n}\left| a_{out,mn}^{+}\pm
a_{out,mn}^{-}\right| ^{2}}
{2\beta _{m}(\left|
a_{in,m}^{+}\right|
^{2}+\left| a_{in,m}^{-}\right| ^{2})},  \label{transm}
\end{equation}
where $a_{out,mn}$ denotes the $m$th mode iuput contribution to the
$n$th mode output $a_{out,n}$
and the sums over $m,n$ is over all possible propagating modes (number
$N_m$)\
in the waveguide. For spin-up electrons injected into a simple
waveguide
of length $l$, we can assume $a_{in,m}^{+}=a_{in,m}^{-}=1/2$ and the
output
coefficients can be calculated using Eq. (\ref{mwg}): 
$a_{out,mn}^{+}=e^{-i%
\alpha _{m}l}\delta_{mn}$ and
$a_{out,mn}^{-}=e^{-i(\alpha _{m}+\Delta )l}\delta_{mn}$.
Since $%
\alpha _{n}=\beta _{n}$  for a simple waveguide, we have $T^{\pm 
}=N_{m}\left|
1\pm e^{-i\Delta l}\right| ^{2}$. The probability of detecting a 
spin-down
${\tiny
\left( 
\array{c}
0 \\ 
1
\endarray
\right)}$
electron will be 
$T^{-}=N_{m}\sin ^{2}(\delta
\theta /2)$ \cite{datt,mire} and that for a spin-up 
${\tiny\left(
\array{c}
1 \\ 
0 \endarray \right)}$ electron $T^{+}=N_{m}\cos ^{2}(\delta \theta 
/2)$, 
where $\delta \theta =\Delta l$ is the phase difference between 
the two spin propagation modes $|+\rangle $ and $|-\rangle $ in the 
structure after traveling a distance $l$.  In stubbed waveguides, it 
will be shown that the percentage of transmitted spin-up and spin-down 
electrons follows that same rule as in a stubless waveguide when
the Rashba effect is weak.  The advantage of stubbed waveguides is 
that using side gates to control the length $h$ of the stubs and the 
distance $d$ of their centers from that of the waveguide, we can 
control the transmission rate.  At the same time we can adjust the 
back gate bias to change the Rashba parameter and then the outgoing 
spin orientation.

Once $T^{\pm}$ is known, the conductance $G^{\pm}\equiv G(E,0)$ at zero temperature is given by
 the Landauer-B\"uttiker formula,
$G^{\pm}=(2e^2/h)T^{\pm}$. For finite temperatures the  conductance $G(E,T)$ is
given by

\begin{equation}
G(E,T)=\int^{\infty}_{-\infty}G(\epsilon,T=0)
\left(-\frac{df}{d\epsilon}\right)d\epsilon ,
\end{equation}
where $f(\epsilon-E)$ is the Fermi function.

In real devices the shape of the stubs can be different than a 
rectangular one
and one question we need to answer is the influence of the shape of 
the stubs on the transmission output.  It is known from  electronic 
stub tuners that stubs of different shape, e.g., Lorentzian or
triangular stubs, do not change the qualitative behavior of the 
transmission.  To quantitatively study the transmission rate of 
electronic current through a waveguide with arbitrarily shaped stub, 
we can break the stubs into a series of rectangular segments 
with the same width $b_i$ and different heights.  Each segment is described 
by a transfer matrix $M_{i}$.  The complete shape is well described by 
the product $M^{T}=\prod_{i}M_{i}$ if the segment width is much 
smaller than the electronic wavelength. The entire procedure described
above for a single stub is then repeated each segment as many times as
required by the particular shape.

\section{RESULTS AND DISCUSSION}

In the numerical calculations we consider a stubbed $ 
In_{0.53}Ga_{0.47}As/InAlAs$ waveguide.  The effective electron mass 
is $ m^{\ast }=0.042m_{0}$ and a typical average spin-orbit constant 
$\alpha =1.6\times 10^{-11}$eVm is assumed throughout the paper 
unless otherwise
specified.  To verify the validity of the perturbation theory we 
evaluated the bound states of one unit,  of stub length $h=2000$\AA\ and 
width $b=150$\AA\, connected to a waveguide of width $a=250$\AA \  and
segment length $l=100$\AA \  to the left and right of the stub.  We
find that the ratio of the intersubband mixing energy over the 
difference between the lowest two bound states is less than $10\%$.  
We also calculated the energy bands of a superlattice of such stub units 
with a waveguide segment of length $l=200$\AA\ between two consecutive 
stubs.  The separation of the two spin bands is less than $15\%$
of the band energy.

\subsection{One mode allowed in the 
waveguide}

\subsubsection{Rectangular stubs}

\begin{figure}[tbp]
\begin{center}
\addvspace{0 cm}
\leavevmode
\epsfysize=250pt 
\epsfxsize=235pt 
\epsffile{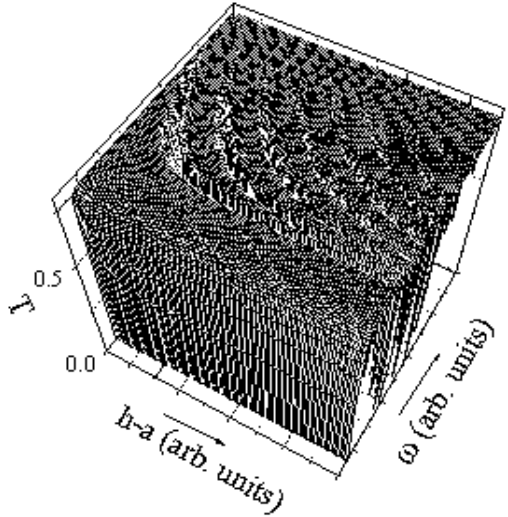}
\end{center}
\caption{Three-dimensional illustration of transmission $T$ versus 
electron energy and stub arm length $h-a$ of a  {\it symmetric} stub tuner.}
\label{fig2}
\end{figure}

To view the parameter dependence of the spin transmission through a 
stubbed waveguide, we inject spin-up polarized electrons into a 
waveguide of width $ a=250$\AA\ with a symmetric double stub of width 
$b=150$\AA\ and measure the output flux rate of spin-up and spin-down 
electrons.  Considering only the first mode and for a waveguide 
segment of length $l=450$\AA, we plot in Fig.  \ref{fig2} the spin-up
electron transmission as a function of the stub length, reduced by 
the width of the waveguide $(h-a)$, and of the electron energy $ E$.  Only 
electrons with energy higher that the first subband in the waveguide 
$E_{1}=\pi ^{2}/(2m^{\ast }a^{2})=14.3$meV can pass through the device 
and the output percentage of spin-up and spin-down electron depends 
only on the total length of the device.  Here $\Delta =2m^{\ast 
}\alpha /\hbar ^{2}=0.1764\times 10^{8}$m$^{-1}$ so the maximum of the 
spin-up electron transmission rate is $\cos ^{2}\Delta (b+l)=76.4\%$ 
and the maximum of the spin-down electron transmission rate $23.6\%$.  The 
stub begins to play a role in adjusting the transmission rate when
electrons have an energy higher than the first subband in the stub 
$E_{1}^{s}=\pi ^{2}/(2m^{\ast }b^{2})=57.2 $meV. If the electronic 
energy has a value between the first and second transverse subbands, 
\textit{i.e.}, if the relation $E_{1}^{s}<E<E_{2}^{s}=2\pi ^{2}/(m^{\ast 
}b^{2})$
holds, 
only one transverse mode enters the stub and a simple transmission 
pattern appears with transmission gaps along the 
curves 
$h=n\lambda +h_{0}$, where $h_{0}$ is the position 
of the first gap and $n=0,1,2,\cdots $.  Notice that the period $\lambda $ is
different from the $x$-direction wavelength of the first subband 
wavefunction in the stub $\lambda _{1}=2\pi b/\sqrt{2m^{\ast 
}Eb^{2}-\pi ^{2}%
}$ and depends on the parameters $a$ and $l$.  When the width $b$ of the stub 
 becomes wider, more modes can exist in the stub and the 
transmission pattern becomes complicated, see Sec. IIIc.

\begin{figure}[tbp]
\begin{center}
\addvspace{0 cm} \leavevmode
\epsfysize=250pt \epsfxsize=235pt \epsffile{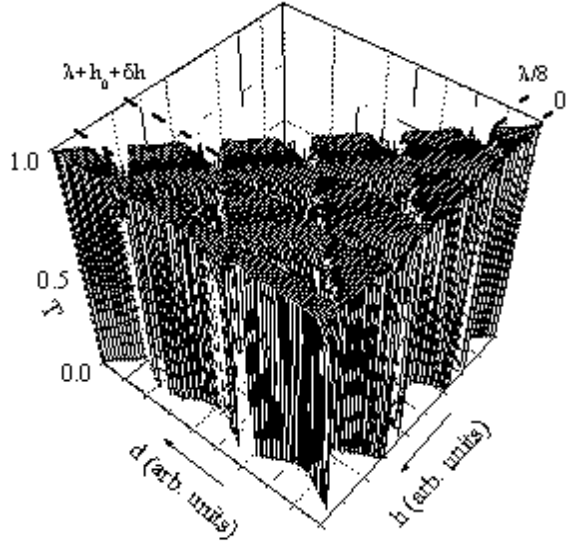}
\end{center}
\caption{Spin-down transmission of an array of  five stub tuners 
as a  function of the stub length $h$\ and stub shift $d$ when only spin-up
electrons are injected. Cross sections along the dashed lines are shown  in Fig. 
(\ref{fig4}).}
\label{fig3}
\end{figure}

We can also change the {\it asymmetry}  parameter $d$, the distance 
along $x$ between the midpoint of the waveguide and that of the 
height $h$ shown in Fig. 1(a), to shift the stub along the $x$ 
direction so the structure becomes  {\it asymmetric}.  In Fig.  
\ref {fig3} we show by a wired surface the transmission rate for 
electrons of fixed energy $E=48$meV as function of $d$ and $h$ for a 
five-stub device with parameters $a=c=250$\AA, $b=150$\AA\ and
$l=207.5$\AA.  The parameter $l$\ is chosen to satisfy $\cos
^{2}[5\Delta (b+l)]=0$ so that spin-up electrons are totally blocked, 
( $T^{+}=0$), and only spin-down electrons come out. 
The transmission is a periodic function of $d$ and $h$ 
and the gaps appear in the triangle-shaped regions.  The centers of 
these triangles are located at the points in the $h-d$ plane 
that satisfy $h=n\lambda 
/2+h_{0}$ and $2d+h=n\lambda +h_{0}$ for integer $n=0,1,2,\cdots $.  
In this figure we find $\lambda =558$ \AA, $h_{0}=694$\AA\ and
$\lambda _{1}=660.5$\AA.  One of the interesting facts found in Fig.
\ref{fig3} is that a square type transmission curve with a wide gap 
can be obtained by changing $d$ for a fixed value of  $h$.

\begin{figure}[tbp]
\begin{center}
\addvspace{2 cm} \leavevmode
\epsfysize=180pt \epsfxsize=250pt \epsffile{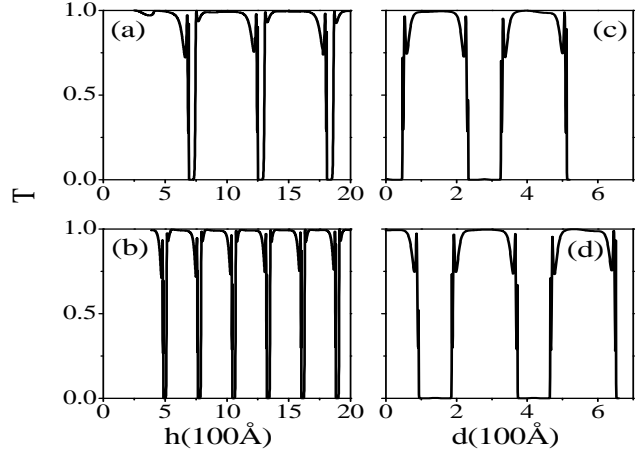}
\end{center}
\caption{Two-dimensional cross section of Fig. (\ref{fig3}) in the
transmission $T$ vs stub length $h$ for fixed $d$ ( $d=0$ in a) and 
$d=\protect%
\lambda /8$ in b) ) and vs stub shift $d$ for fixed $h$ ( 
h=1289.5\AA\ in c) and h=1568.5\AA\ in d) ).}
\label{fig4}
\end{figure}

\noindent To clearly show when and how these square-shaped 
transmission curves appear, some cross sections of Fig.  \ref{fig3} 
are shown in Fig.  \ref {fig4}.  In Fig.  \ref{fig4}(a) the spin-down
transmission rate is shown as a  function of $h$ for the symmetric case 
($d=0$).  Gaps appear with period $%
\lambda $ and the first gap begins at $h=h_{0}=694$\AA\ and has a 
width $%
\delta h=40$\AA.  If we cut the surface  along
$h$\ at $%
d=69.75$\AA \ $=\lambda /8$, we obtain Fig.  \ref{fig4}(b).  Here we
see a periodic structure similar to that in (a) but 
with half the period $\lambda /2$ and half the gap width $\delta 
h=20$\AA.  If we fix the value of the stub length $h$\ in one of the 
gaps in Fig.  \ref{fig4}(a) and cut the surface of Fig.  \ref{fig3}
along $d$, we obtain the curve shown  in (c), where 
$h=1289.5$\AA \ $\simeq h_{0}+\lambda $.  The transmission has a period
$\lambda /2$ along $d$\ with gaps as wide as $\delta d=91$\AA.  The
same periodic transmission versus $d$ curves is shown (d) 
when we make the same cut as above but fix $h=1568.5$\AA \
$\simeq h_{0}+3\lambda /2$ so the gaps appear at $d=\lambda /4$ rather 
than at $d=0$.  We see that in (c) and (d) we  can 
completely flip or block the input electronic spin flux and the 
transmission has an almost square-wave dependence on the adjustable 
parameter $d$ with the gap/band ratio as big as 0.5.

\subsubsection{Stubs of general shape}

\begin{figure}[tbp]
\begin{center}
\addvspace{2 cm} \leavevmode
\epsfysize=150pt \epsfxsize=235pt \epsffile{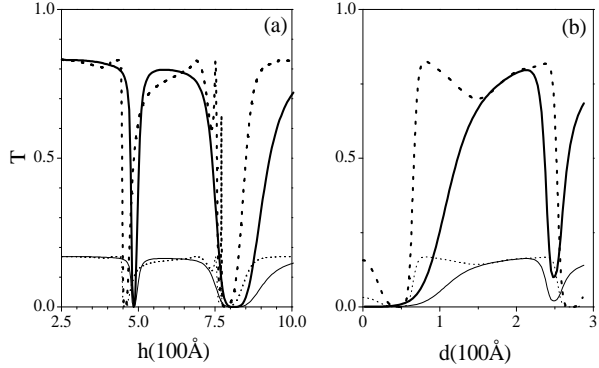}
\end{center}
\caption{Transmission as a function of the stub 
length $h$, for symmetric double stubs (a), and of the stub shift $d$ for 
asymmetric double stubs (b). The thin and thick solid curves show,  
respectively, 
the spin-up and spin-down transmission for Gaussian shaped stubs. The
 thin and thick  dotted curves show the corresponding results for rectangular 
stubs
  with parameters  $a=250$\AA, $b=290$\AA, $L=360$\AA.} 
\label{fig5}
\end{figure}

To study the influence of the stub shape  on  
the transmission output we proceed as outlined at the end of Sec.
II. As an example we consider devices 
 with double stubs of Gaussian shape.  As a function of $x$ the $y$ coordinate 
of the boundary of each stub is taken as  $y=(h-a)/2\pm 
 d](e^{-8x^{2}/b^{2}}-e^{-2})/(1-e^{-2})$.  In Fig.  \ref{fig5}(a) we
 show the transmission of a structure of two symmetric double stubs as 
 a function of the total length $h$ of the stubs.  Only spin-up 
 electrons are injected.  The thin and thick solid curves show, respectively, 
the spin-up and spin-down transmission for Gaussian-shaped stubs and
 parameters $a=250$\AA, $b=375$\AA, $l=275$\AA;   the thin and thick 
 dotted curves  are for   rectangular  with parameters
 $a=250$\AA, $b=290$\AA, $L=360$\AA. We see that the transmission 
 through the waveguide with Guassian-shaped stubs is similar to that with 
 narrower   rectangular  stubs for the same total length ($b+L$).  
 Such  a similarity was noticed earlier between rectangular  and 
 triangular stubs \cite{wang}.   If we fix the total length of the stubs of both waveguides $h=825$\AA \
 and shift the stubs, we obtain Fig.  (\ref{fig5} b)) for  the transmission 
 as a function of the shift   $d$.

\noindent

\subsection{Two modes allowed in the waveguide}

\begin{figure}[tbp]
\begin{center}
\addvspace{1 cm} \leavevmode
\epsfysize=280pt \epsfxsize=175pt \epsffile{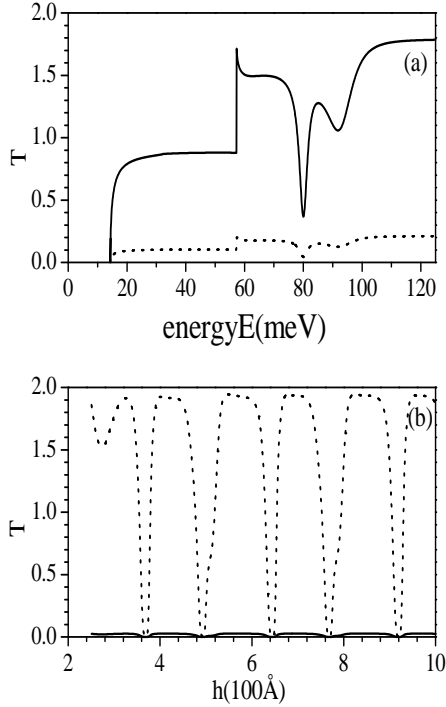}
\end{center}
\caption{(a) Spin-up $T^{+}$ (solid line) and spin-down
 $T^{-}$ (dotted line) transmission through   one  stub tuner  as
a function of the electron energy $E$. (b) $T^{+}$\ and $T^{-}$  for  
four identical stub tuners  as a function of the  stub length $h$.}
\label{fig6}
\end{figure}

When the electron energy is high enough, two or more transverse
modes can propagate in the waveguide.  In Fig.  \ref{fig6}(a),
we plot the transmission as a function of the electronic energy $E$ through 
a device with one waveguide segment of width $a=250$\AA\ and length 
$l=250$\AA\ and one stub of width $b=125$\AA\ and length $h=1250$\AA.
Here the total device length is adjusted so that both spin-up (solid line) 
and spin-down (dotted line) electrons can be observed at the right 
end.  When the electron energy is lower than $E_{2}=57.2$meV, only one 
propagating mode exists and the transmission   is almost flat;  the total 
transmission ($ T^{+}+T^{-}$) 
approaches  unity because in this energy region the stub is too narrow
to give significant contribution.  With the increase of the electron 
energy, two modes can exist in the structure.  In some cases, two 
modes compete  and a deep minimum exists with the transmission dropping from 2 
to a value much lower than 1 as shown in Fig.  \ref{fig6}(a).  If a
wider stub is used here, more modes can exist in the stub and the 
curve has many more oscillations.

In Fig.  \ref{fig6}(b) we inject electrons of energy $100$meV into a
device of four equal units, each of which has one asymmetric stub with 
$b=125$\AA\ and $d=30$\AA\ connected to one waveguide segment of width 
$a=250$\AA\ and length $l=300$.  The transmission rate is shown as a  function 
of the 
length $h$ of the stubs.  In this figure the spin-orbit 
parameter is adjusted to $\alpha =1.8\times 10^{-11}$eVm so that 
almost all electrons are spin-flipped after passing the device.  As 
in Fig. 4 we  observe a series of gaps, narrower than the corresponding ones of 
Fig. 4,  but now with  two modes present  and electrons 
of either spin completely blocked. Notice that the transmission rate  is still a 
periodic function of the 
stub length $h$ despite the presence of  two 
modes.

\subsection{Introducing defects}

\begin{figure}[tbp]
\begin{center}
\addvspace{2 cm}
\leavevmode
\epsfysize=150pt \epsfxsize=215pt \epsffile{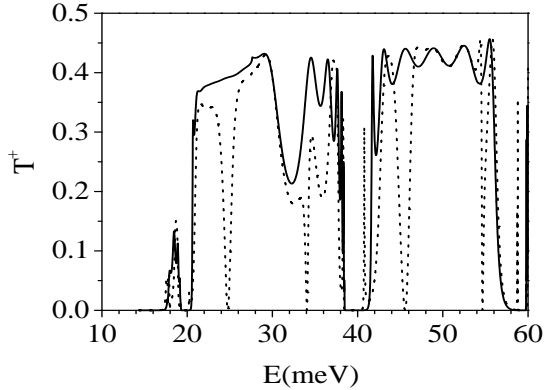}
\end{center}
\caption{Spin-up transmission $T^{+}$ versus electron energy $E$ 
 for a periodic array of seven units 
(solid line) and for the same structure with
the central (fourth) unit replaced by a defect unit.}
\label{fig7}
\end{figure}

Using a series of periodically arranged stub units, we can make a 
spintronic crystal (superlattice) with a transmission similar to that 
shown in Fig.  \ref{fig3}, 
where five units are used.  In this kind of crystals, we can expect to 
obtain devices of better behavior than that of only one unit.  If we 
add some impurities into the crystal, we can change the properties of 
this crystal and observe a different transmission rate.   The solid line 
in Fig.  
\ref{fig7} shows the first mode spin-up transmission 
rate versus electron energy through a pure finite crystal of seven 
equal units with symmetric stubs and parameters $a=250$ \AA, $b=250$\AA,
 $l=125$\AA,  and $h=2250$\AA.  The transmission shows gaps near
$E=10$meV, $40$meV and $60$meV. When we substitute the 
middle (fourth) unit by a defect, that is, a unit  which has stub length 
$h_{d}=4500$\AA
 and otherwise the same parameters, we find that the defect 
introduces resonant peaks in the gaps at $E=40$ meV and $E=46$ meV and 
$E=55$ meV. The spin-down transmission rate has the same structure but 
different scale. A similar behavior was reported for photonic tuners 
in Ref. \cite{tak2} 

\subsection{Injection of spin-up and spin-down electrons}

\begin{figure}[tbp]
\begin{center}
\addvspace{2.5 cm} \leavevmode
\epsfysize=150pt \epsfxsize=295pt \epsffile{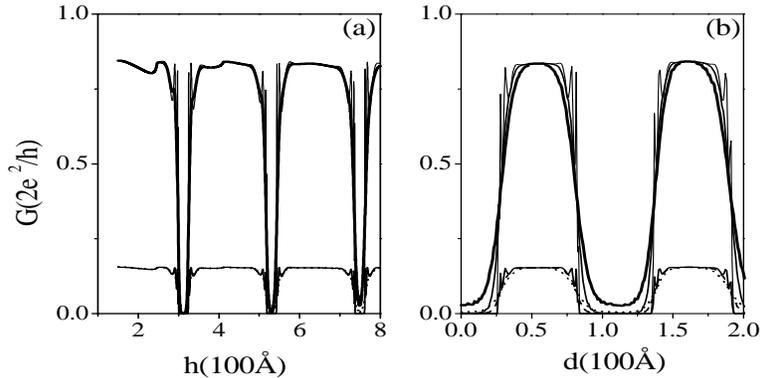}
\end{center}
\caption{Spin-up $G^{+}$ (lower lines) and spin-down
conductance $G^-$ (upper lines) as functions  of the stub length $h$ (a) 
and of the stub shift $d$ (b). The various curves and temperatures 
are specified in the text.}
\label{fig8}
\end{figure}

Up to now we considered electronic conductance or transmission
through stubbed 
waveguides when only 
spin-up electrons are injected at zero temperature ($G(E,T=0)$).
In this subsection we discuss what happens if arbitrarily 
spin-polarized electrons are injected into a stubbed waveguide at finite
temperature.  In
the $\sigma_z$ spin representation the spin-orientation is denoted by 
the coefficients $C^+$ and $C^-$. For instance,   
${\tiny \left( \array{c} 1 \\
\pm 1
\endarray
\right)}$
 describes a  spin oriented along the $x$-direction whereas  
${\tiny \left(
\array{c}
1 \\
\pm i \endarray \right)}$
 describes a spin oriented along the $y$-direction.  In Fig.  
\ref{fig8}(a) and (b) we show the spin-up (dotted line) and spin-down
(solid line) conductance for a structure composed of eight stub units
with $a=150$\AA, $b=80$\AA, $l=69$\AA\  and Fermi energy $E_F=133$meV.
The spin orientation of the incident
electrons is as follows:  we let $C^+=0.7$ and $C^-=-0.3$; since the state must
be normalized ($|C^+|^2+|C^-|^2=1$), this corresponds to
$84.5\%$   spin-up   and $15.5\%$ spin-down incident electrons.
As a function of the stub length $h$ the
conductances $G^+$ and $G^-$, at $T=$0 and 8K, are
 shown in Fig. \ref{fig8}(a);
 the solid and dotted curves are for  $G^+$ and
the thin and thick solid lines for  $G^-$. As a function of the shift $d$
$G^+$ and $G^-$ are shown in Fig. \ref{fig8}(b)  at $T=$0, 4.2, and 8K.
The solid, dashed, and dotted curves are for $G^+$ and the
thin, medium, and thick solid lines for $G^-$.  
Here we have $15.5\%$ of the output electrons having  spin up and $84.5\%$
of them  spin down. The conductance $G^-$ shows a  behavior similar to that of the transmission
in Fig.  \ref{fig4}(a) and (d). 
 As can be seen, finite temperatures
smoothen the curves obtained at zero temperature similar to the
case of spinless electrons \cite{tak1}. As the ratio $E_F/k_BT$
increases the curves are smoothened or rounded off more strongly. 
 
\section{Concluding remarks}

Employing the transfer-matrix method  we  combined the spin precession in a 
waveguide, due to
the spin-orbit coupling, with the basic physics of a stub tuner, and applied
it to the ballistic spin transmission through periodically stubbed waveguides.
We found that the spin polarization can be well controlled by adjusting the
total length of the device for a wide range of the electronic energy. 
In particular, we showed that, given  spin-polarized electrons injected into a
stub structure, we can select the spin of the outgoing electrons to be the same
as or opposite to that of the injected electrons. 
We demonstrated this for stubs of rectangular or Gaussian shape but
also for triangular stubs \cite{wang}. In general these results hold
for stubs of any shape. The latter does not affect the qualitative behavior of
the transmission but only its period when several stubs are combined.

More important, we saw clearly that, as a function of the
stub height $h$ and the asymmetry parameter $d$, we can have a
nearly {\it binary square-wave transmission } (spin-valve effect) for either 
spin orientation,
with wide gaps for stubs of different shapes and a
well-controlled range of the stub parameters. In this respect {\it asymmetric}
stubs give the best results. Their asymmetry, i. e., $h$ and $d$,  
can be controlled by lateral gates \cite{debr};  in principle, such gates 
should allow for a more detailed control of the overall stub shape. 
A further modulation can be  achieved if we 
combine several groups of stubs with different values of the spin-orbit coupling
strength. These findings should facilitate the experimental 
realization of the spin transistor.

A qualitative understanding of all these results is as follows.
The spins precess in a single waveguide \cite{datt} due to the spin-orbit 
coupling.
On the other hand, in a stub tuner  waves reflected from the walls of the stub,
where the wave function vanishes, may interfere constructively or destructively
with those propagating in the main waveguide and result, respectively, in an 
increase  or
decrease of the transmission \cite{sols}. Refining this idea,  it was shown in 
Refs. 
\cite{debr}-\cite{tak1} that using double stubs the transmission of {\it 
spinless}
electrons could be blocked completely using {\it asymmetric} stubs. Combining
several stubs would result in a nearly square-wave transmission output, 
especially
as a function of the asymmetry parameter $d$. The transmission shown, e.g.
 in  Figs. \ref{fig2}-\ref{fig4} is simply the result of  this behavior  when 
combined with the spin
precession due to the spin-orbit coupling since the length of the device
was  chosen such that spin flip would occur in the stubless waveguide.

The most clear results or simplest transmission patterns are obtained
when only one mode is allowed to propagate in the main waveguide. 
If more  modes can propagate in the main
waveguide, generally the transmission pattern becomes more complex
or even irregular. However, as we demonstrated, we can have a simple periodic 
transmission pattern
even when two modes are allowed. This occurs when the stub width is short
enough that only one mode enters in the stub region.

Further, we showed that  the above results hold when the injected spins 
are polarized in an arbitrary direction, partially 'up' and partially 'down'.
We also showed that introducing 'defects' 
in a finite superlattice leads to a further modulation of the spin current 
since the 'defects' produce new transmission resonances 
or antiresonances. One could  use such 'defects' to achieve a
specific control of the transmission. 
 
Finally, we have seen that  the effect of finite temperatures
is to smoothen the zero-temperature conductance as in  the
case of spinless electrons \cite{tak1}. The degree of  
smoothness depends mainly on the ratio $E_F/k_BT$.

\section{Acknowledgement}
 The work of X. F. W. and  P. V. was supported by the  Canadian NSERC Grant No.
OGP0121756  and that of F. M. P. by the
 Flemish Science Foundation, the Belgian Interuniversity Attraction
 Poles (IUAP), the Inter-university Microelectronics Center (IMEC) and
 the Concerted action programme (GOA).

\end{document}